# Two-dimensional material based "field-effect CCD"

Hongwei Guo[1,2,#], Wei Li[2,#], Jianhang Lv[2], Akeel Qadir[2], Ayaz Ali[2], Lixiang Liu[2], Wei Liu[2], Yiwei Sun[2], Khurram Shehzad[2], Bin Yu[3], Tawfique Hasan[4] & Yang Xu[2,5,*]

[1]Centre for Optical and Electromagnetic Research, State Key Laboratory for Modern Optical Instrumentation, Zhejiang Provincial Key Laboratory for Sensing Technologies, Zhejiang University, Hangzhou, 310058, China.

[2]School of Information Science and Electronic Engineering, College of Microelectronics, Zhejiang University, Hangzhou, 310027, China.

[3]College of Nanoscale Science and Engineering, State University of New York, New York, 12203, U.S.A.

[4]Cambridge Graphene Centre, University of Cambridge, Cambridge, CB3 0FA, UK.

[5]State Key Laboratory of Silicon Materials College of Material Science, Zhejiang University, Hangzhou, 310027, China.

[*]Email: yangxu-isee@zju.edu.cn

[#]These authors contributed equally to this work.

**Abstract**

**Charge-coupled device (CCD), along with the complementary metal-oxide-semiconductor (CMOS), is one of the major imaging technologies. Traditional CCDs rely on the charge transfer between potential wells, which gives them advantages of simple pixel structure, high sensitivity, and low noise. However, the serial transfer requires fabrication incompatible with CMOS, and leads to slower, more complex, and less flexible readout than random access. Here, we report a new-concept CCD called field-effect CCD (FE-CCD), which is based on the capacitive "coupling" between the semiconductor substrate and the 2D material. The strong field effect of the 2D material enables amplification, and non-destructive readout of the integrated charges in one pixel without transfer. We demonstrated the broadband response of our CCD (up to 1870 nm) and the successful integration of 2D hetero-structures for reducing the power consumption. The demonstrated FE-CCD provides a valuable strategy for versatile 2D materials to be monolithically integrated into the semiconductor imaging**

**technology.**

**Introduction**

Charge-coupled device (CCD)[1] and complementary metal-oxide-semiconductor (CMOS) are two main imaging technologies[2-6]. With enormous efforts towards higher performance and larger pixel density, both technologies keep competitive and are widely used in high-end applications and commercial products. For CCD, the simple MOS photogate pixel and charge transfer structure give rise to high sensitivity, high fill factor (FF), and low noise. But high charge transfer efficiency (CTE) requires specialized fabrication process which hinders the CMOS-compatible monolithic integration[4,7]. The serial transfer also requires relatively complicated multi-phase biasing/clocking which could limit the readout speed and increase the power consumption[2]. For CMOS, independent pixel structure allows random access, simple clocking, high-speed parallel readout, natural anti-blooming, and low power consumption[4]. CMOS also has the advantage of monolithic integration with multiple functionalities such as readout and data processing circuitry[2-5]. On the other hand, CMOS usually has lower FF and higher noise than CCD due to the relatively complex active pixel sensor (APS)[4] structure and subsequent circuitry. Thus, imaging devices combining CCD's MOS photogate and CMOS's independent pixel structure can have significant advantages in monolithic integration, performance and readout[8,9].

Despite of many achievements in CCD and CMOS, silicon-based imager has difficulties in working beyond the visible range, which is usually overcome by the non-monolithic integration of narrow-bandgap bulk semiconductors[10,11]. 2D materials not only show enormous potential in monolithic integration with CMOS[12,13], but also have wide choices of the bandgap[14] to obtain the broadband response. By incorporating the 2D materials, we can enhance both the responsivity and the spectral performance of Si-based photodetector or imager[12,15-19]. However, the typical photodiode[15,16] or photoconductor[17-19] structure based on 2D materials and Si are difficult to achieve high responsivity and high linearity at the same time, which is a bottleneck for practical imaging applications[20].

In this work, we reported a novel detecting/imaging device concept called field-effect CCD (FE-CCD), which is based on CCD's MOS photogate but requires no charge transfer between pixels (*i.e.* "couple"). The "couple" is re-defined as the capacitive coupling[21,22] between the semiconductor substrate and the 2D material (*e.g.* graphene). In the semiconductor, we created the potential well for charge integration by applying a gate voltage pulse. In the 2D material, the non-destructive and direct readout along with the charge signal amplification was realized by the strong field effect. Besides, the charge integration in our FE-CCD is beneficial for the low-light-level condition, and gives high linearity for accurate image capturing. The FE-CCD also shows a broadband response from visible to short-wavelength infrared (SWIR) wavelength, and its power consumption is readily suppressed by using the 2D-material hetero-junction.

**Results**

**Charge integration and readout in the FE-CCD.** Our FE-CCD is based on the 2D-material field effect transistor (FET), but requires a lowly-doped semiconductor substrate. Fig. 1a shows the schematic of our FE-CCD pixel, which consists of a simple Gr/SiO$_2$/*n*-Si (GOS) capacitor and can be fabricated in few steps (see Methods). As in the typical CCD, we created the potential well (deep depletion[21]) in the FE-CCD by applying a fast-sweeping or pulsed gate voltage $V_g$. The photo- and thermally generated holes are integrated in the well, while the corresponding electrons are transferred to graphene through the outer circuit (see Supplementary). The deep depletion was confirmed by the continuous decrease of the capacitance for $V_g > 5$ V in the high-frequency capacitance-voltage (HF-CV, 100 kHz) of the GOS capacitor, when $V_g$ was swept at relatively fast rate of 10 V/s from accumulation to inversion (Fig. 1b). The photo-hole integration is also demonstrated in the HF-CV by the increase of the capacitance with laser power. In order to examine the effect of deep depletion and charge integration on the transfer characteristics of graphene, we measured the drain current $I_d$ as a function of $V_g$ ($I_d$ - $V_g$) of our FE-CCD at the same voltage sweep rate (Fig. 1c). A strong correspondence between the HF-CV and the $I_d$ - $V_g$ was observed, as the $I_d$ - $V_g$ also shows a large photo-response (see Supplementary) in the deep-depletion regime ($V_g > 5$ V). The $I_d$ decreases with the increasing power, because the number of the photo-electrons transferred to *p*-type graphene increases, as monitored simultaneously by the gate charging current $I_g$ (Fig. 1e). Derivative of the transfer characteristics gives the quasi-static capacitance-voltage (QS-CV, Fig 1d), confirming that the variation of the drain current can exactly measure the charge variation in the GOS capacitor (see Supplementary). Thus, compared to the serial transfer and readout in traditional CCDs, the direct and non-destructive readout in our FE-CCD is made possible by the strong field effect of graphene.

To demonstrate the practical operation of our FE-CCD, we applied the gate voltage pulse, and obtained the periodic drain current waveform ($I_d$ - $t$) in dark and light conditions (Fig. 1f). The variation of the hole number in the well is monitored in real time by the variation of the $I_d$ - $t$. The FE-CCD integrates the holes in the pulse region ($V_g$ = 40 V), and clears them in the reset region ($V_g$ = 0 V). In the pulse region, the dark $I_d$ - $t$ keeps almost constant since the thermal generation requires relatively long time to fill the well (storage time[23] ~ 80 s, see Supplementary). The light $I_d$ - $t$ first decreases (linear region) and then saturates (saturation region) within relatively short time (saturation time ~ 0.5 s), indicating that the well is filled by the photo-holes. This large difference of drain current between the light and dark condition results from the integration effect and the signal amplification (photo-conductive gain[24]) from the graphene, both of which can give high sensitivity to our FE-CCD.

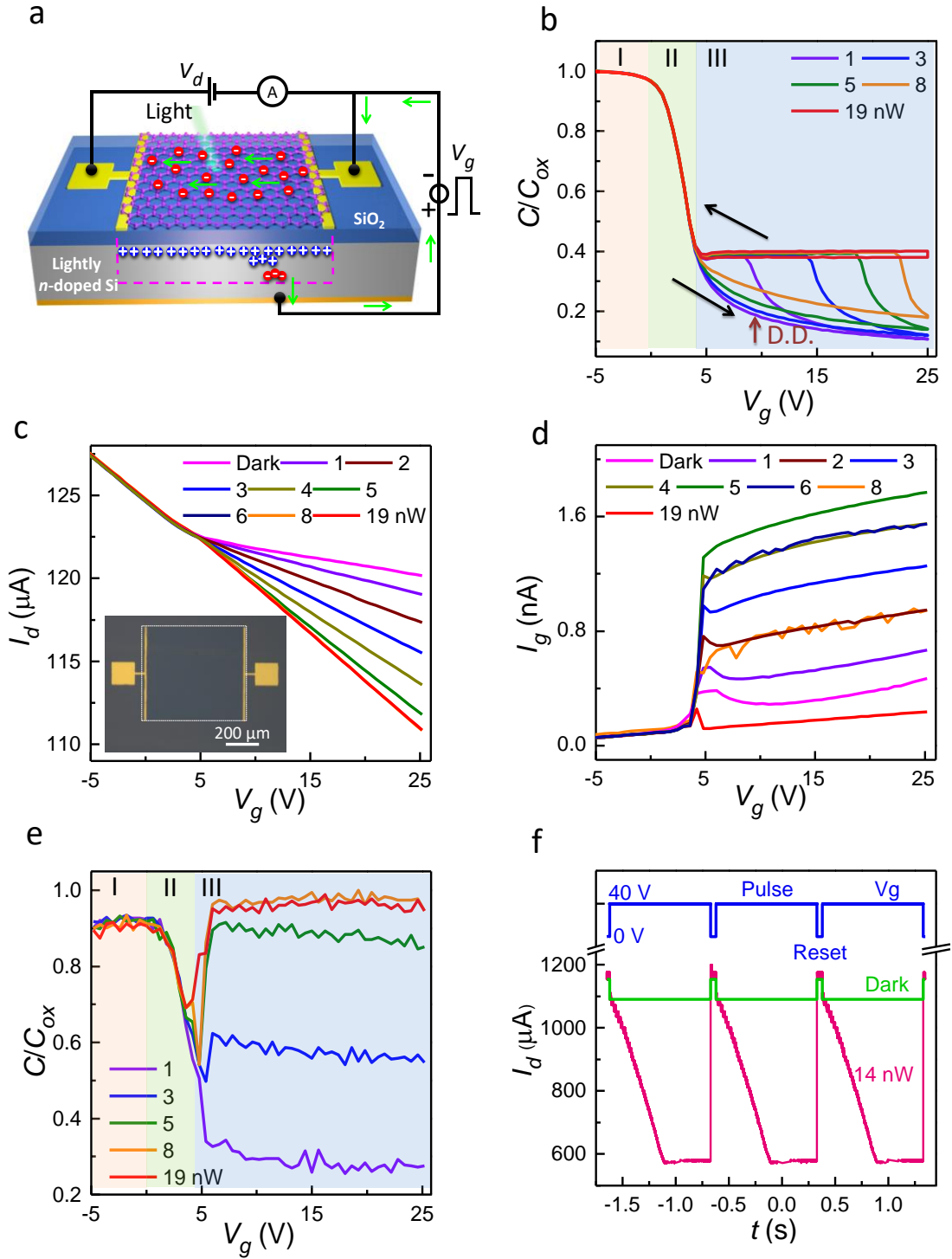

**Figure 1.** (a) Schematic of our FE-CCD pixel. The gate can be driven by fast-sweeping or pulsed voltage. Holes (blue spheres) are generated and integrated in the potential well with corresponding electrons (red spheres) transferred to graphene. (b) The normalized HF-CV (100 kHz) of the FE-CCD at different laser power (wavelength 532 nm). Gate voltage (defined to be positive on Si) is first swept from negative voltage to positive voltage and then swept back at rate of 10 V/s. Doping concentration of the $n$-type Si: $\sim 3\times10^{15}$ cm$^{-3}$. Area of graphene and pads: 0.35 mm$^2$. Oxide thickness:

100 nm. Oxide capacitance $C_{ox}$: 120 pF. Region I, II, and III represent the accumulation, depletion, and inversion regime, respectively (same below). Deep depletion is labelled as "D. D.". (c) Transfer characteristics $I_d$ - $V_g$ ($V_d$ = 0.1 V, $p$-type branch) of the FE-CCD at different laser power (wavelength 532 nm). The gate voltage is swept from negative voltage to positive voltage at rate of 10 V/s. Inset: Optical image of the FE-CCD. Scale bar: 200 μm. (d) The gate charging current $I_g$ - $V_g$ monitored simultaneously with $I_d$ - $V_g$. (e) The normalized QS-CV at different power derived from $I_d$ - $V_g$. (f) Typical drain current waveform ($I_d$ - $t$) of our FE-CCD in dark and light (14 nW) condition under periodic gate voltage pulses. The drain voltage $V_d$ is 1.5 V (same for all $I_d$ - $t$ measurements).

**Mechanism and performance of the FE-CCD.** To understand the detailed integration process, we studied the power-dependent $I_d$ - $t$ of our FE-CCD within one integration period (Fig. 2a). As shown by the fitting, the experimental $I_d$ - $t$ can be described by a linear-saturation model (see Supplementary), and the corresponding photo-hole number is deduced (right axis). This linearity of the $I_d$ - $t$ before the saturation is important for accurate image sensing as the photo-hole number (exposure dose) is proportional to the product of time and light intensity. The slope of the $I_d$ - $t$ can give the photo-electron charging current $I_g$ (see Supplementary), which characterizes the light intensity as in the photodiode (Fig. 2b). The fitting of $I_g$ - $P$ gives $R^2$ of 0.9990, indicating a high linearity of our FE-CCD. From the fitting, we also obtained relatively high quantum efficiency $\eta$ of (76 ± 1) %, which can be attributed to the high transparency of the graphene[25]. Besides the constant illumination, the integration process is also demonstrated with laser pulses, as shown by the step-like waveform of the $I_d$ - $t$ (Fig. 2c). This multi-pulse integration shows that our FE-CCD can potentially work as a typical multi-bit photo-memory as well[26]. The photo-hole integration process mainly happens within the FE-CCD device region, as verified by the spatial photocurrent mapping (SPCM) at different integration times (Fig. 2d).

The performance of our FE-CCD can be characterized by the similar parameters that used in traditional CCDs. Compared to the state-of-art conversion gain (CG) of ~ 10 μV/e⁻ and FWC of ~ 100 ke⁻[20], the present FE-CCD gives relatively low CG of 5 pV/e⁻ and high FWC of $10^8$ ke⁻ due to the non-optimized large pixel area of 0.25 mm² (assuming $I_d$ is converted to voltage with resistive gain of $10^3$ Ω). Theoretically, by improving the graphene mobility ($\mu$) to $10^5$ cm²/(V·s)[27] and down-scaling the pixel to 10 μm², CG can reach as high as 160 μV/e⁻, while FWC maintains a large value of 108 ke$^-$. The CG ($\propto \mu V_d/L^2$, $V_d$ the drain voltage, $L$ the pixel length) can be independently improved through the mobility and the drain voltage without affecting FWC, which is difficult to achieve in conventional CCDs due to the floating-diffusion capacitance trade-off[20] (see Supplementary). The FWC can be increased by increasing the gate voltage, as shown by the decrease of saturation current with $V_g$ (Fig. 2e). The noise equivalent voltage (NEV) and the corresponding number of noise electrons for the present FE-CCD is 5.57 μV Hz$^{-1/2}$ and 2.5×10$^5$ ke$^-$, respectively (see Supplementary). The linear dynamic range (LDR) of 52 dB was subsequently deduced by FWC ($V_g$ = 40 V) over the number of noise electrons.

The performance of our FE-CCD is also strongly affected by the background dark current, we thus studied the temperature-dependent $I_d$ - $t$ in dark condition (Fig. 2f). For relatively low temperature, the dark $I_d$ varies slowly with time, indicating relatively slow thermal generation. For relatively high temperature, the thermal generation is fast enough to fill the potential well, as shown by the saturation in the $I_d$ - $t$. The Arrhenius plot of $I_g$ gives the activation energy of 0.8 eV and 1.1 eV (~ Si bandgap) for $T < 400$ K and $T \geq 400$ K, respectively, indicating that the generation mechanism is temperature-dependent and changes from the depletion-dominated generation to diffusion-dominated generation[28]. This large thermal generation rate at high temperature impedes the integration of the photo-holes. Thus, the FE-CCD requires to work at relatively low temperature to obtain the high-quality photo-signal.

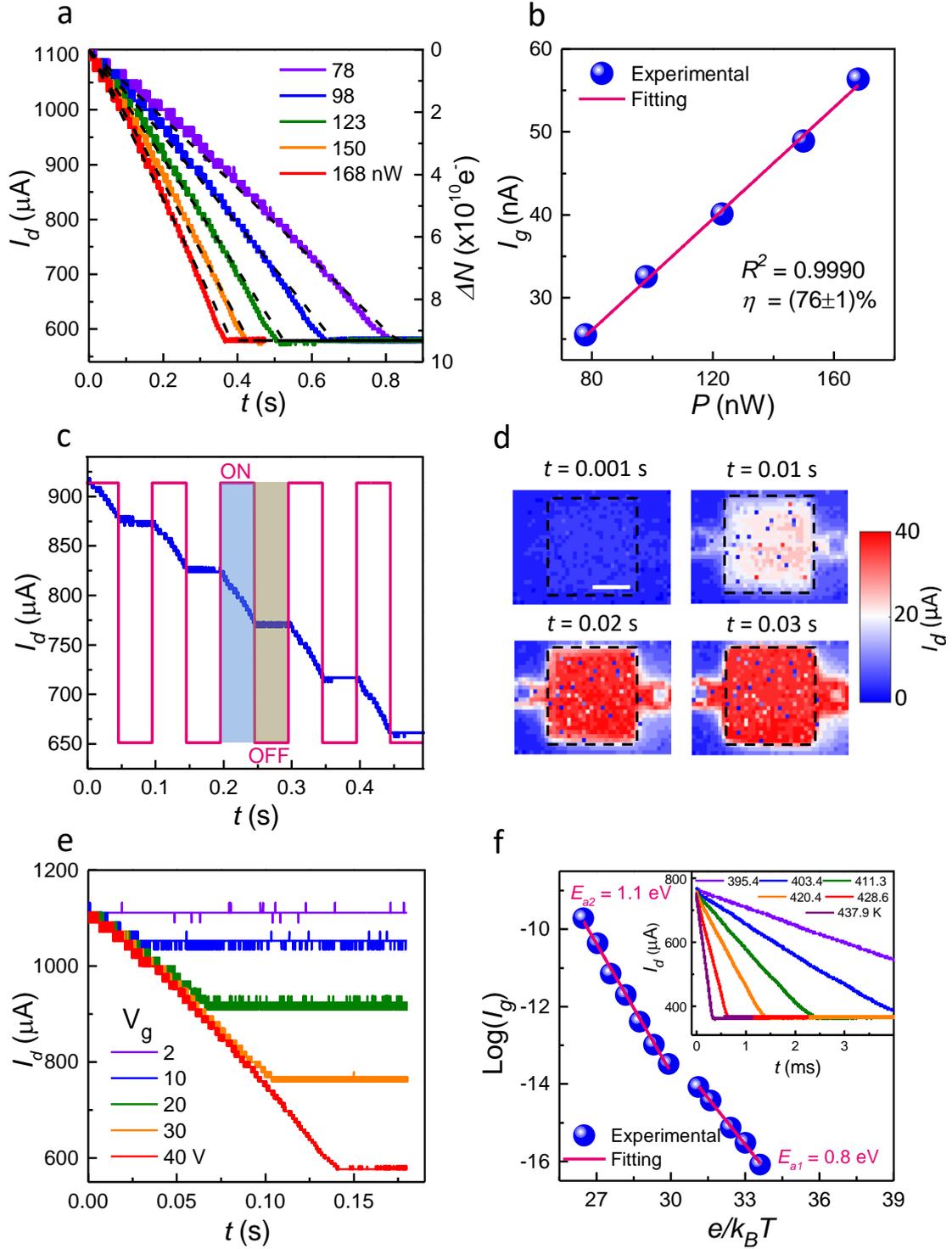

**Figure 2.** (a) $I_d$ - $t$ (solid) of the FE-CCD at different power for the wavelength of 532 nm. Only one period is shown (same below). The fitting (dash) is based on the linear-saturation model. The variation of charge number is also shown on the right axis. $V_g$ = 40 V. (b) The gate charging current $I_g$ as a function of power $P$, obtained from the slope of the $I_d$ - $t$ linear region the in (a). The linear fitting gives the QE of (76 ± 1) %. (c) $I_d$ - $t$ of the FE-CCD at the pulsed laser condition (532 nm, 132 nW). The laser pulse was synchronized with the gate pulse. $V_g$ = 50 V. (d) SPCM of the FE-

CCD obtained at different integration time by using the laser spot (532 nm, 2.5 µW) with diameter of 40 µm. In the SPCM, the photocurrent was obtained through a 1 kΩ resistor in series with the graphene channel, and the initial background current was subtracted. $V_g$ = 20 V. The dotted square represents the graphene region. Scale bar: 200 µm. Photo-response in the region apart from the graphene is due to the scattered light entering the potential well through the edges of electrode. (e) $I_d$ - $t$ of the FE-CCD at different gate voltage in light condition (532 nm, 360 nW). (f) Arrhenius plot of $I_g$ extracted from the dark $I_d$ - $t$ (inset) at different temperature. The activation energies for thermal generation are 0.8 eV and 1.1 eV (~ Si bandgap) for 300 ~ 400K and 400 ~ 450 K, respectively. $V_g$ = 50 V.

**SWIR response of the FE-CCD.** Our FE-CCD has a broadband response with wavelength extended to 1870 nm (see Supplementary), which is important for SWIR imaging[12]. Fig. 3a shows a representative power-dependent $I_d$ - $t$ at the wavelength of 1342 nm. As shown in the $I_d$ - $t$, our FE-CCD gives a much larger photo-response (*e.g.* $I_d$ variation of ~ 200 µA at 3.04 mW) than the typical III/V IR photodetector (responsivity ~ 1 mA/W[29]). To figure out the origin of the SWIR response, we performed the SPCM on our device. Photo-response was observed not only in the whole graphene-covered region, but also in the graphene-free region (edges of pads), which indicates that the generation is probably from the Si as the generation from the graphene should only happen at the metal-graphene junction due to the fast Auger recombination[30]. The generation from the Si is further confirmed by SWIR response in the HF-CV of the semi-transparent Au/SiO$_2$/Si control device (Fig. 3b, see Supplementary). We thus attributed this SWIR response (sub-bandgap generation) to the surface-state absorption (SSA) which provides the middle-gap levels at the SiO$_2$/Si interface[31] (inset). The quantum efficiency for the sub-bandgap generation (~ 10$^{-5}$) is usually much lower than that for the visible generation, as shown by the $\eta$ spectrum obtained from the $I_d$ - $t$ of FE-CCD (Fig. 3c). Thus, the photo-hole integration in the well and the strong field-effect of the graphene can largely enhance the photo-response even for the low quantum efficiency, and this photo-response can be further improved by increasing the FWC and photo-conductive gain through $V_g$ and $V_d$, respectively.

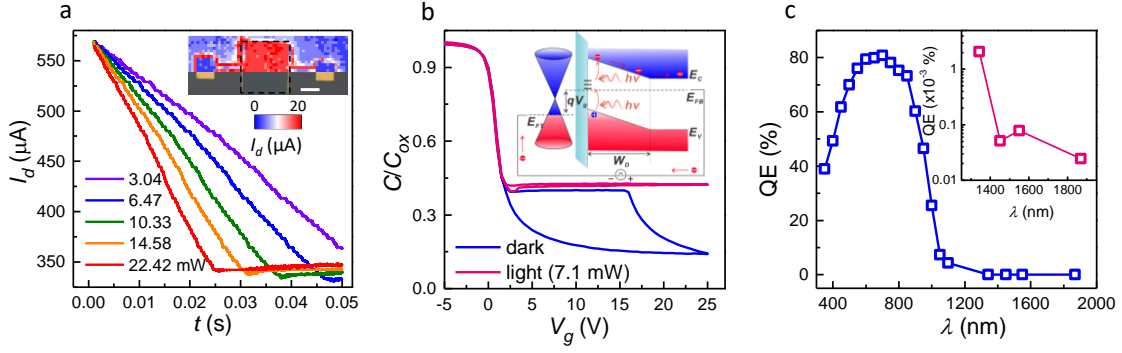

**Figure 3.** (a) $I_d$ - $t$ of the FE-CCD at different power for the wavelength of 1342 nm. $V_g$ = 50 V. Inset: The device optical image, and the SPCM obtained at the integration time of 2 s by using the laser spot (1342 nm, 1 mW) with diameter of 10 μm. Scale bar: 200 μm. (b) The normalized HF-CV (100 kHz) of the semi-transparent Au (10 nm)/SiO$_2$/Si control device in dark and light (1342 nm, 7.1 mW) condition. Inset: Band diagram showing the SSA process in the GOS capacitor. $W_D$ is the width of the deep-depletion region. $E_C$, $E_V$, $E_{FB}$, and $E_{FT}$ represent the conduction band minimum, valence band maximum, bottom Fermi level (Si), and top Fermi level (Gr), respectively. (c) The spectrum of the QE in visible and SWIR range. Inset: The QE in the SWIR range.

**Two-dimensional hetero-junction based FE-CCD (FE-JCCD).** Due to the zero bandgap, the graphene-based FE-CCD has relatively large background dark current, which requires high power consumption (~ 1 mW for the above devices). We can use 2D materials with moderate bandgap (*e.g.*, transition metal dichalcogenide, TMDC[14]) or their hetero-junctions, which can be truly turned "OFF" to suppress the dark current. Fig. 4a shows the schematic and the optical image of the Gr/WSe$_2$/Gr FE-JCCD, in which WSe$_2$ and graphene work as lateral channel and contact, respectively (see Methods). To study the channel "ON/OFF" characteristics, we first measured the transfer characteristics of our FE-JCCD at relatively fast voltage sweep (Fig. 4b). The WSe$_2$ is initially lowly *n*-doped and in "OFF" state with $I_d$ less than 1 nA, as observed from the drain current at $V_g$ = 0 V. In dark condition, the deep-depletion region continuously expands as the thermal generation cannot catch up with the voltage sweep (Fig. 4c, top band diagram). Accordingly, relatively few electrons are transferred to the WSe$_2$, and most of the $V_g$ drops on the deep-depletion region, giving ineffective gating and hence the channel remains in the "OFF" state. In light condition, the photo-generation in the potential well can give more electrons to the WSe$_2$ channel (Fig. 4c, bottom band diagram) and turn the channel "ON", resulting in a high light-to-dark "ON/OFF" ratio of ~ 10$^4$. We then obtained the $I_d$ - $t$ at different power to confirm this dark current suppression, as shown in Fig. 4d. Corresponding to the transfer characteristics, the $I_d$ - $t$ in the dark is much lower (down to the noise level) than in the light. The light current shows a threshold time which decreases with the increasing power. This threshold time is determined by the critical number of the electrons required to turn "ON" the channel. When the channel is in "ON" state, we can read out the photo-charges through the $I_d$ - $t$. Though the linearity might be lower than the graphene case, the background power consumption is readily reduced to the 1 nW level.

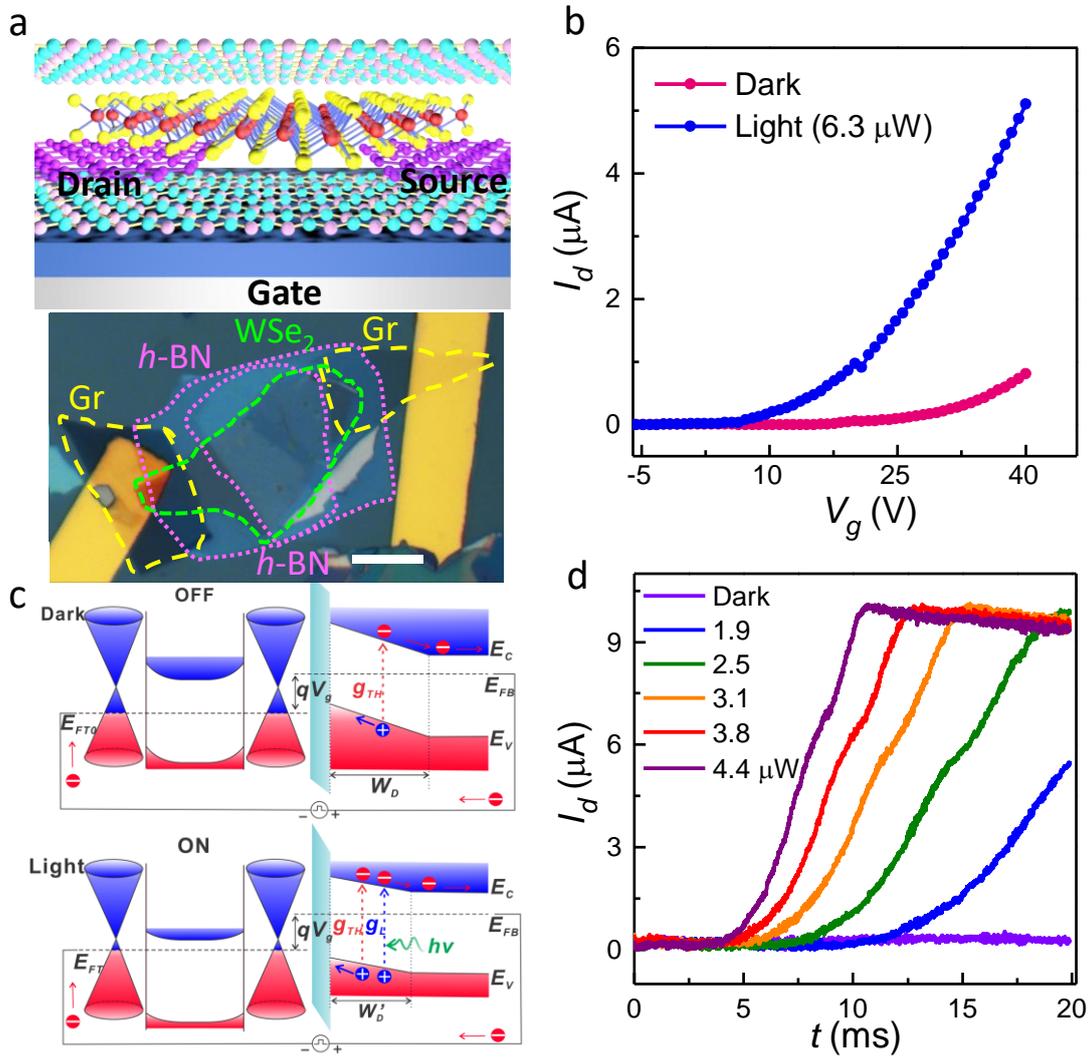

**Figure 4.** (a) Schematic (top) and optical image (bottom) of the Gr/WSe$_2$/Gr FE-JCCD by van der Waals hetero-stacking. The Gr/WSe$_2$/Gr is encapsulated by the $h$-BN. Scale bar: 10 μm. (b) Transfer characteristics of the FE-JCCD in dark and light (532 nm, 6.3 μW) condition. The sweeping rate of the gate voltage is 16 V/s. $V_d$ = 2.5 V. (c) Band diagrams of the FE-JCCD in the dark ("OFF", top) and light ("ON", bottom) condition. $g_{TH}$ and $g_L$ represent the thermal generation and the photo-

generation, respectively. (d) $I_d$ - $t$ of the FE-JCCD at different power for the wavelength of 532 nm.

**Imaging and charge transfer based on FE-CCD.** To demonstrate the application of image capturing, we performed the two-dimensional scanning of the single graphene-based FE-CCD pixel, and obtained the reflection images (size 160×160) under incandescent light (Fig. 5(b, c)). The intensity and contrast of the images increase with the integration time as the exposure dose increases. The relatively long integration time is thus important in improving the image quality, especially for the low-light-level condition. Though our single pixel device demonstrates the successful imaging, large pixel array is required for practical imaging applications. Recent developments in large scale and high-quality chemical-vapor-deposition (CVD) synthesis of graphene and related materials[32] enable the fabrication of the pixel arrays of our FE-CCD. In Fig. 5a, we demonstrated the proof-of-concept FE-CCD linear array, which is able to work in both random access mode and charge transfer mode (results not shown here).

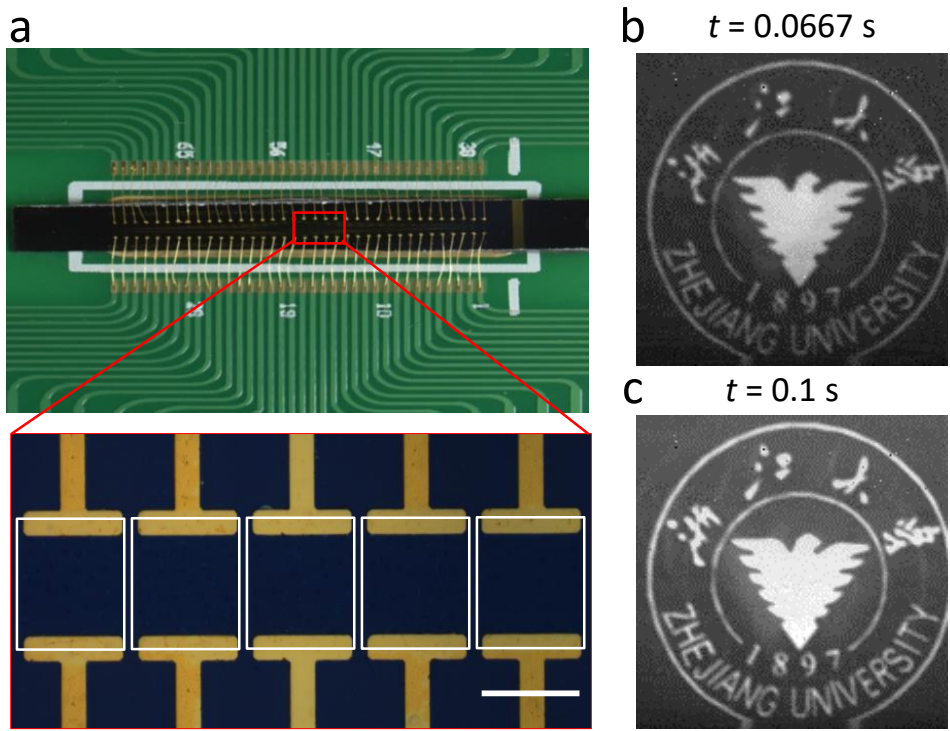

**Figure 5.** (a) The proof-of-concept FE-CCD linear array working in both random-access mode and charge transfer mode. Top: the FE-CCD linear array wire-bonded to the printed-circuit board. Bottom: the enlarged the optical image of the FE-CCD linear array. Scale bar: 50 μm. (b, c) Reflection images at different integration time obtained by scanning the single graphene-based FE-

CCD pixel under the incandescent light (power density $5\times10^{-5}$ W/cm$^2$).

**Discussion**

The imaging performance of our FE-CCD is determined by not only the semiconductor substrate, but also the 2D material. On one hand, the bandgap and the surface states of the semiconductor substrate are two important parameters that determine the working wavelength and the dark thermal generation of the FE-CCD, respectively. The Si has a typical cut-off wavelength of ~ 1100 nm (Fig. 3c), and the wafer we used gives the dark current density of ~ 80 nA/cm$^2$ (considering the FWC of $10^{11}$ e$^-$ and the storage time of 80 s), which is much larger than the typical dark current density of 0.1 ~ 1 nA/cm$^2$ for the buried-channel CCD (BCCD)[33]. This relatively large dark current is probably due to the thermal generation through the surface states, similar to the case of surface-channel CCD (SCCD)[34]. On the other hand, the choice of the 2D material along with its mobility and doping determines the output characteristics of the FE-CCD. Graphene has relatively high mobility, which can dramatically improve the CG, as the drain current is proportional to the mobility. For the doping, the drain current of the highly *p*-doped CVD graphene decreases with the increasing number of transferred photo-electrons, while the drain current of the lowly *p*-doped graphene from mechanical exfoliation shows a non-monotonic behavior, as the Fermi level in the lowly *p*-doped case is much easier to be tuned over the Dirac point (see Supplementary). Thus, considering the linearity of the charge-to-current conversion, the doping type should remain the same during the transfer of photo-charges. For the 2D semiconductor such as the mechanically exfoliated WSe$_2$, though the doping maintains *n*-type, the linearity is not as good as the CVD graphene with the drain current showing a switch behavior due to the bandgap, and the relatively low mobility of the WSe$_2$ is one of the main reasons for the small drain current in the FE-JCCD.

Besides the semiconductor substrate, the 2D material can also contribute to the photo-response. One contribution can be the photo-conductive response from the 2D material itself (see Supplementary). This photo-conductive component can be attributed to the traps within the 2D material or the unintentional contamination[35]. The other contribution can be from the photo-charges directly generated within the 2D material[36,37]. These photo-charges can also be transferred to the semiconductor and then integrated, since the 2D material in the MOS capacitor plays similar roles as the semiconductor substrate. Thus, by incorporating 2D materials with different bandgap (*e.g.*

black phosphorous, TMDC)[38], the FE-CCD can provide a large potential in the wide-spectrum response from the long-wavelength infrared (even THz) to the ultra-violet.

In conclusion, we demonstrated a new-concept FE-CCD for photo-detecting and imaging. In the FE-CCD, a potential well is created by driving the lowly doped MOS capacitor into non-equilibrium deep depletion through the gate voltage pulse. The photo-charges are linearly integrated in the potential well, while the corresponding opposite charges are transferred to the 2D material which enables the direct readout through the field effect. The FE-CCD can thus be randomly accessed, and requires no charge transfer between the potential wells, as demonstrated by the scanning image from a single FE-CCD pixel. The FE-CCD shows a broadband response including visible and SWIR (1100~1870 nm) range, and the relatively large SWIR response is attributed to the amplification of the photo-charges generated from SSA process. The successful integration of the 2D van der Waals hetero-structure (Gr/WSe$_2$/Gr) into our FE-CCD suppressed the background power consumption to 1 nW level. We finally demonstrated the charge transfer between wells in the FE-CCD linear array, in which all the charge states can be monitored in real time. Based on the results of our study, we propose that we can further improve or tailor the performance of the FE-CCD by implementing other 2D materials or hetero-structures. Furthermore, even in the traditional imaging technology, the highly transparent 2D materials can be monolithically integrated as the gate material, offering the flexible charge readout and high performance.

**Methods**

**Device fabrication.** We first evaporated the Au/Cr (60 nm/5 nm) electrode onto the lowly doped (1-10 Ω·cm) *n*-type SiO$_2$/Si (100 nm/500 μm) wafer. For the graphene-based FE-CCD (including linear array), we transferred the CVD graphene onto the electrodes using polymethyl methacrylate (PMMA)[18] as a supporting layer, and then etched the graphene into desired pixel shape by O$_2$ plasma. For the Gr/WSe$_2$/Gr FE-JCCD, we mechanically exfoliated, stacked and transferred the junction onto electrodes in all-dry process[27]. We then wire-bonded the devices to the printed-circuit board.

**Electrical and opto-electronic measurements.** In this work, the light sources are single lasers (532 nm, 1342 nm, 1450 nm, 1550 nm, and 1870 nm), incandescent lamp, and Xe lamp with a color filter (300 ~ 1200 nm). The power of all light sources was calibrated by the commercial Si (Thorlabs

S120VC) and InGaAs photodetector (Thorlabs S184C). The current-voltage and capacitance-voltage were measured by Agilent B1500A. For the pulse test, we used a two-channel signal generator (RIGOL DG1000Z) and a voltage amplifier (Falco Systems WMA-300) to generate large-voltage pulses. The output current from 2D materials was measured by the trans-impedance amplifier (Femto HSA-Y-2-40) connected to the oscilloscope (Lecroy WJ354A). For the temperature-dependent $I_d$ - $t$ test, we used the hot plate to increase the temperature. For the pulsed laser test, the laser was modulated by the signal generator and synchronized with the gate voltage pulse. For the SPCM and photograph, we built a scanning system with focused laser and programmable X-Y stage. The data acquisition (DAQ) card was used to obtain the $I_d$ - $t$ data. For the spectrum measurement, we calculated QE from the $I_d$ - $t$ under Xe lamp and independent lasers.

**Acknowledgements**

This work is supported by ZJ-NSF (LZ17F040001), Fundamental Research Funds for the Central Universities (2016XZZX001-05, 2017XZZX008-06, 2017XZZX009-02), National Science Foundation China (NSFC, Nos. 61674127 and 61474099), and the National Key Research and Development Program of China (2016YFA0200204, 2016YFA0301204).